\documentclass[referee]{raa}            

\usepackage{graphicx,times}             
\usepackage{amssymb,amsmath}
\usepackage{natbib}

\begin{document}

   \title{Photometric and spectroscopic analysis of LBV candidate J004341.84+411112.0 in M31}

   \volnopage{Vol.0 (200x) No.0, 000--000}      
   \setcounter{page}{1}          

   \author{A. Sarkisyan
   			\inst{1}
   		   \and O. Sholukhova
   		   	\inst{1} 
   		   \and S. Fabrika
   		   	\inst{1} 
   		   \and A. Valeev
   		   	\inst{1,4} 
   		   \and A. Valcheva
   		   	\inst{2}
   		   \and P. Nedialkov
   		   	\inst{2}
   		   \and A. Tatarnikov
   		   	\inst{3}	
     }

   \institute{
   	Special Astrophysical Observatory of Russian Academy of Sciences, 
Niznij~Arkhyz 369167, Karachai-Cherkessia, Russia; {\it ars@sao.ru}\\
	\and
	Department of astronomy, Sofia University, 5 J. Bourchier blvd, Sofia 
1164, Bulgaria   
	\and
    Sternberg Astronomical Institute, Moscow M. V. Lomonosov State University, Universitetskiy prospekt~13, Moscow 119234, Russia
	\and
	Crimean Astrophysical Observatory, Russian Academy of Sciences, Nauchnyi 298409, Russia
	}

   \date{Received~~2021~~month day; accepted~~2021~~month day}

\abstract{
	We study Luminous Blue Variable (LBV) candidate J004341.84+411112.0 in the Andromeda galaxy. 
    We present optical spectra of the object obtained with the 6-m telescope of SAO RAS. 
	The candidate shows typical LBV features in its spectra: broad and strong hydrogen lines and the \ion{He}{i} lines with P Cigni profiles. Its remarkable spectral resemblance to the well known LBV P Cygni suggests a common nature of the objects and supports LBV classification of J004341.84+411112.0. 
    We estimate the temperature, reddening, radius and luminosity of the star using its spectral energy distribution. 
	Obtained bolometric luminosity of the candidate ($M_{\text{bol}}=-10.41\pm0.12$\,mag) is quite similar to those of known LBV stars in the Andromeda galaxy.
	We analysed ten year light curve of the object in \textit{R} filter.
	The candidate demonstrates photometric variations of the order of 0.4\,mag, with an overall brightness increasing trend $\Delta$\textit{R}\,$>$\,0.1\,mag. Therewith, the corresponding colour variations of the object are fully consistent with LBV behavior when a star become cooler and brighter in the optical spectral range with a nearly constant bolometric luminosity. LBV-type variability of the object, similarity of its spectrum and estimated luminosity to those of known LBVs allows us to classify J004341.84+411112.0 as a LBV. 
}

   \authorrunning{A. Sarkisyan et al.}            
   \titlerunning{LBV candidate J004341.84 in M31}  

   \maketitle

%
%

\section{Introduction}           

In the present paper we report results of optical spectral and photometric monitoring of LBV candidate in the Andromeda galaxy. We select object J004341.84+411112.0 from the list of LBV candidates by \citet{Massey2007,Massey2016a} for this study. The star was firstly discovered as LBV candidate by \citet{Massey2006b} and was named by author as ``P Cygni analog in M31'' for its unique spectral similarity to one of the archetypes of LBVs, P Cygni. However, despite remarkable spectral resemblance to P Cygni, author could not establish variability of the object and had to consider it as LBV candidate. 
LBVs show various spectral features during the course of their life \citep{Humphreys2014} and different type of brightness variation on the different time scales \citep{vanGenderen2001} which makes it difficult to classify them. One of the main properties of LBV is the presence of S Dor-type variability when a star become cooler and brighter in the optical spectral range with a nearly constant bolometric luminosity \citep{HumphreysDavidson1994}. Based on this, we try to prove that the object J004341.84+411112.0 belongs to the class of LBV stars. Earlier we have already studied J004341.84+411112.0 as a part of work on selected LBV candidates in M31 \citep{Sarkisyan2020}. Here we continue and further our investigation with the help of extended long-term spectral and photometric observations. Below we analyze ten year light curve of the object, describe its spectra, construct spectral energy distribution (SED) and use it to determine the stellar parameters for the consequent classification. We will refer to the star by its first (RA) coordinate throughout the paper. 

\section{Observations}

The optical spectra of the J004341.84 were obtained with the SCORPIO spectrograph \citep{AfanMois2005} on the 6-m telescope BTA SAO RAS in October 2012, October 2019, September 2020 and October 2021. The observing log is shown in Table~\ref{tab:spec}. The spectra have been reduced using standard procedures for long-slit spectroscopy. They were bias and cosmic ray subtracted, co-added over multiple exposures, corrected for flat-field, wavelength and flux calibrated. The spectra extraction were performed with the \textsc{spextra} code designed to work with stars in crowded fields \citep{Sarkisyan2017}.

Several photometric sets were also obtained with the SCORPIO: in addition to the spectra simultaneous photometry the object was also taken in January 2015 and September 2016. Magnitudes of the object were estimated by comparison with nearby stars in the field, with the help of Local Group Galaxies Survey \citep[LGGS,][]{Massey2016b} calibration. 
Results of the optical photometry with the SCORPIO are given in Table~\ref{tab:phot}.

The additional optical photometric data of the object have been acquired with the 50/70 cm Schmidt telescope at NAO Rozen, Bulgaria which was equipped with FLI 4096x4096\,pix CCD camera with a scale of 1.07\,arcsec\,pix$^{-1}$. The observations cover a period of 3\,years (2016/07/09 - 2019/07/08) and contain mainly \textit{R} and \textit{B} band images and only 7 epochs in \textit{V} band. The seeing conditions were between 2 and 4\,arcsec with a typical value of 2.5\,arcsec. The raw images were darkfield subtracted and flatfielded using standard \textsc{iraf} \citep{Tody1986,Tody1993} routines. The images of a given night in each band were aligned and combined (usually 3x300 or 5x300\,s) and an aperture photometry of the object of interest and the standard calibration stars was done by \textsc{iraf/daophot} package \citep{Stetson1987,Stetson1990} with aperture radius around FWHM of the image. We used LGGS catalog \citep{Massey2016b} for standard calibration of our instrumental magnitudes.

Near-infrared (NIR) photometric observations of the object were carried out using ASTRONIRCAM camera \citep{Nadjip2017} attached to 2.5m telescope at the Caucasus Mountain Observatory of Sternberg Astronomical Institute \citep{Shatsky2020}. Observations were performed using the dithering method, with the telescope shifting between frames by 5\,arcsec. All frames were corrected for non-linearity, dark current and flat field. For \textit{JHK} photometry, we used summary images with effective expositions of 780, 840 and 930\,sec, respectively. The stars 2MASS J00433931+4111319 and J00433889+4111202 were used as a comparison stars for calibration. Their brightness were taken from 2MASS catalogue \citep{Cutri2012} and converted into MKO photometric system \citep{Simons2002,Tokunaga2002,Tokunaga2005} using the colour equations from \citet{Leggett2006}.

\begin{table}
	\begin{center}
		\caption{The spectroscopic observations log. 
			The columns represent the date of the observations, average seeing, exposition time, used grism, spectral ranges and resolution as full width at half maximum (FWHM)}\label{tab:spec}		
		\begin{tabular}{cccccc}
			\hline\hline
			Date         & Seeing   & Exposition & Grism & Spectral range & Resolution \\
			(DD.MM.YYYY) & (arcsec) & time (s)   &       &     (\AA)      & (FWHM, \AA) \\
			\hline
			15.10.2012 & 1.4 & 2x900  & VPHG550G  & 3500--7200 & 11  \\
			16.10.2012 & 1.1 & 2x900  & VPHG1200G & 4000--5700 & 5.3 \\
			25.10.2019 & 1.5 & 2x1200 & VPHG1200B & 3600--5400 & 5.5 \\
			25.10.2019 & 1.5 & 2x600  & VPHG1200R & 5700--7500 & 5.3 \\
			28.09.2020 & 1.1 & 2x600  & VPHG550G  & 3500--7200 & 11  \\
			28.09.2020 & 1.1 & 2x600  & VPHG1200G & 4000--5700 & 5.3 \\
			02.10.2021 & 2.2 & 4x900  & VPHG1200B & 3600--5400 & 5.5 \\
			02.10.2021 & 2.2 & 3x900  & VPHG1200R & 5700--7500 & 5.3 \\			
			\hline
		\end{tabular}
	\end{center}
\end{table}
\begin{table}
    \begin{center}
	    \caption{The optical photometry with the SCORPIO and NIR photometry with ASTRONIRCAM. We show the date of the observations and estimated magnitudes with their uncertainties.}\label{tab:phot}
	
	    \begin{tabular}{c*{5}{|c}}
		\hline\hline 
		Date & \textit{U} & \textit{B} & \textit{V} & \textit{R} & \textit{I} \\ 
		\hline 
        16.10.2012 &                & 18.10$\pm$0.05 & 17.56$\pm$0.05 & 17.17$\pm$0.05 &                 \\
        17.01.2015 &                & 17.98$\pm$0.08 & 17.48$\pm$0.08 & 17.02$\pm$0.05 & 16.81$\pm$0.05  \\
        26.09.2016 &                & 17.80$\pm$0.05 & 17.35$\pm$0.05 & 16.88$\pm$0.05 &                 \\
 		25.10.2019 &                & 17.92$\pm$0.05 & 17.40$\pm$0.05 & 17.03$\pm$0.04 &                 \\
 		28.09.2020 & 17.45$\pm$0.08 & 17.87$\pm$0.09 & 17.44$\pm$0.04 & 17.01$\pm$0.04 & 16.79$\pm$ 0.09 \\ 
 		02.10.2021 & 17.44$\pm$0.10 & 17.89$\pm$0.06 & 17.34$\pm$0.04 & 17.93$\pm$0.03 & 16.62$\pm$ 0.12 \\ 
 		\hline 
     	 & & \textit{J}$_{\text{MKO}}$ & \textit{H}$_{\text{MKO}}$ & \textit{K}$_{\text{MKO}}$ &  \\ 
     	\hline
        18.12.2018 & & 16.32$\pm$0.05	& 16.09$\pm$0.05 & 15.80$\pm$0.05  & \\
 		\hline
	    \end{tabular}
	\end{center}    
\end{table}

\section{Spectra}
\label{sect:spec}

\begin{figure}
	\includegraphics[width=0.98\columnwidth]{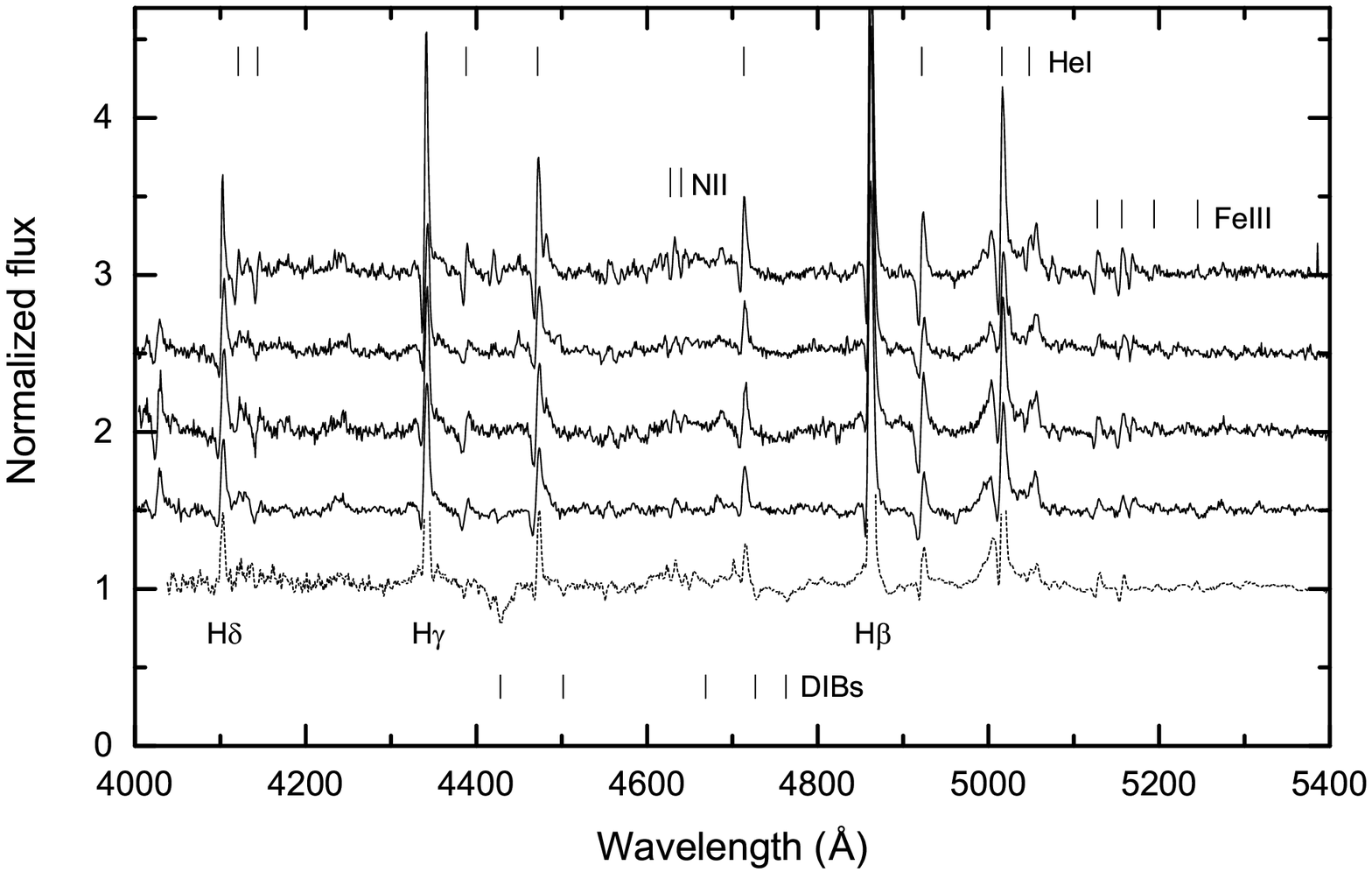}\\
	\includegraphics[width=0.98\columnwidth]{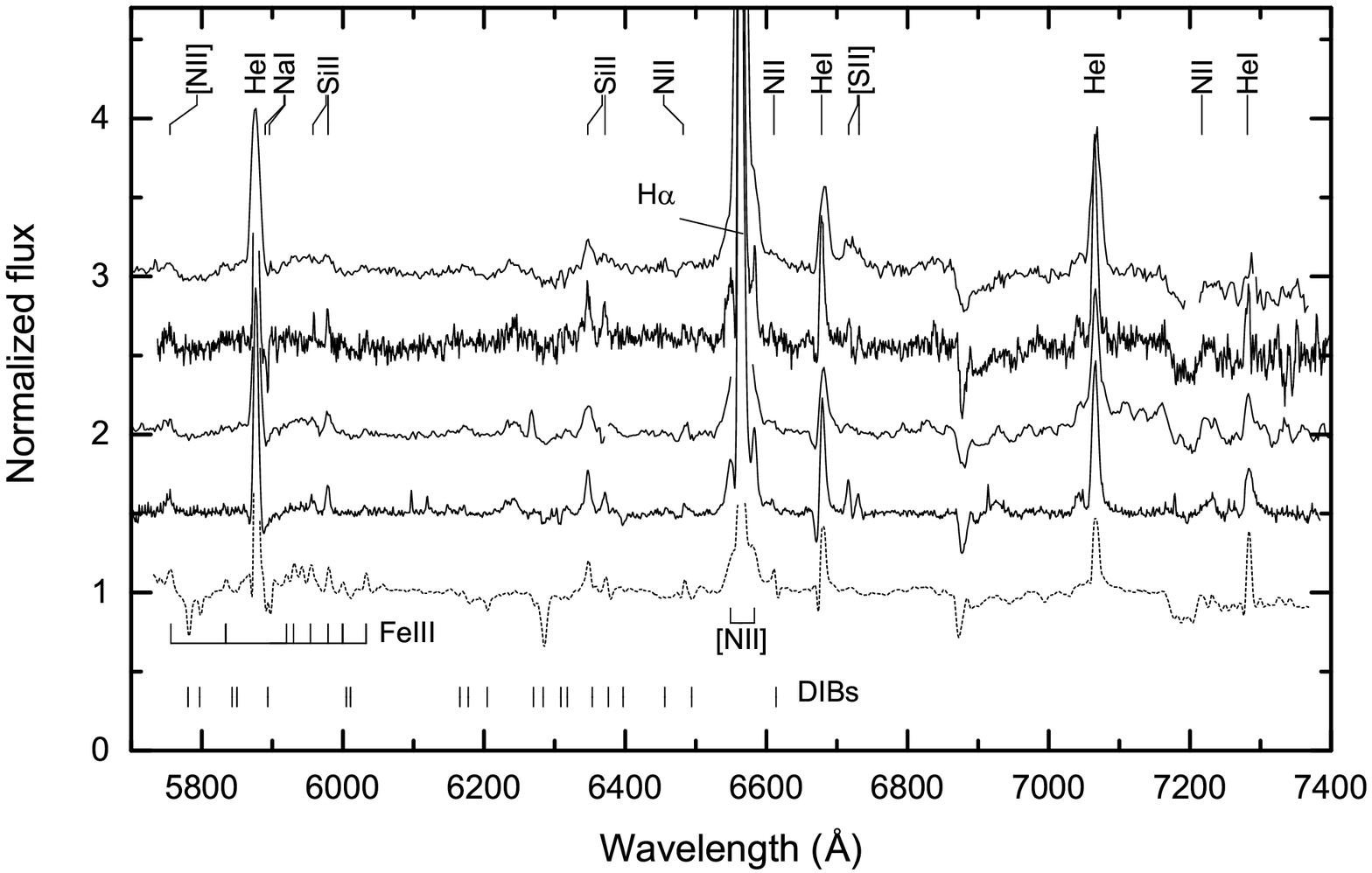} 
	\caption{The optical spectra of the J004341.84 and MN112. The solid lines represent spectra of the J004341.84 for October 2012, October 2019, September 2020, October 2021 epochs (from top to bottom). The spectrum of the MN112 from \citet{Sarkisyan2020} is shown with a dotted line for comparison. The principal strong lines and diffuse interstellar bands (DIBs) are identified.}\label{fig:specs_op}
\end{figure}

Figure~\ref{fig:specs_op} demonstrate optical the spectra of our star. In addition to the spectrum of 2012 from \citep{Sarkisyan2020}, here we present the spectra of 2019, 2020 and 2021. 

We also show spectrum of the Galaxy LBV candidate MN112 from \citet{Sarkisyan2020}. We used MN112 spectrum for comparison since it is almost identical to the spectrum of the well known LBV P Cygni \citep{Gvaramadze2010,Kostenkov2020}.
Figure~\ref{fig:specs_op} clear reveals that the all spectra of J004341.84 is extremely
similar to the spectrum of MN112 and consequently, to that one of P Cygni. One can directly compare our spectra with the spectrum of P Cygni from \citet{Stahl1993}: the spectrum of J004341.84, like P Cygni, is dominated by strong emission lines of hydrogen and \ion{He}{i}, the Balmer lines have wide wings and weak P Cygni absorption components, while the \ion{He}{i} and \ion{Fe}{iii} lines show strong P Cygni profiles, numerous emission lines of \ion{Fe}{iii}, \ion{Si}{ii}, \ion{N}{ii} are also present.

So both the stars J004341.84 and MN112 have P Cygni profiles in hydrogen, \ion{He}{i} and \ion{Fe}{iii} lines, and their spectroscopic similarity with the bona fide LBV, P Cygni, is evident. This remarkable spectral resemblance of the star to P Cygni suggests a common nature of the objects, which is decent argument in support of LBV classification of J004341.84.

\begin{figure}
	\begin{center}
	\includegraphics[width=0.5\columnwidth]{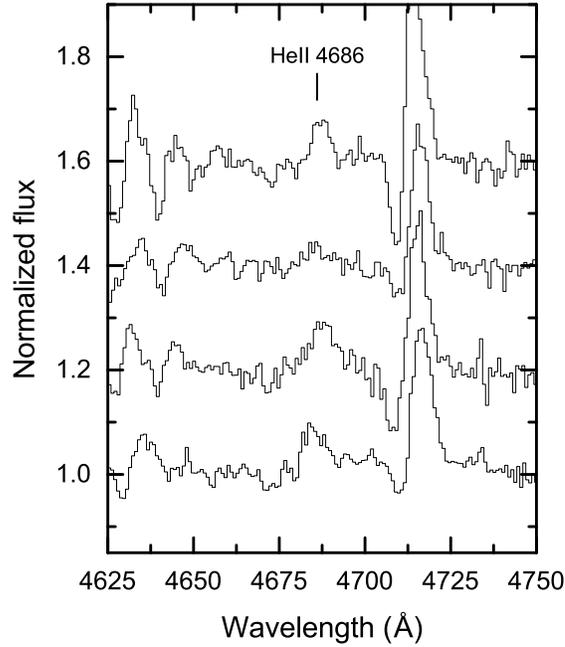}
	\caption{The optical spectra of the J004341.84 in the region of \ion{He}{II} 4686 line. The order of the spectra is the same as in figure~\ref{fig:specs_op}. The \ion{He}{II} 4686 line significantly weakened from 2012 to 2019 and then strengthened again from 2019 to 2020 and 2021.}\label{fig:specs_he}
	\end{center}
\end{figure}

Our spectra confirm previous classification of the object as an Of/late-WN star \citep{Massey2007,Humphreys2014}. However, we would like to draw attention to the following features of our spectra: P Cyg profiles of hydrogen lines, lack of \ion{N}{III} 4634-40-42 lines, characteristic for Of/late-WN star, and quite weak \ion{He}{II} 4686 line, the one of distinctive lines for Of/late-WN stars. Moreover, during our monitoring we have noted the variation of \ion{He}{II} 4686 line. Figure~\ref{fig:specs_he} demonstrate the significant weakening of the \ion{He}{II} 4686 line from 2012 to 2019 and the reverse strengthening from 2019 to 2020 and 2021. Similar disappearing \ion{He}{II} 4686 line were demonstrated by two Of/late-WN stars R127 \citep{Stahl1983} and HDE 269582 \citep{Stahl1986a}, that subsequently have been classified as LBVs \citep{Walborn2017}. So we tend to think that this behavior of the object point to its LBV instability.

\section{Spectral energy distribution}
\label{sect:sed}

For our LBV candidate J004341.84 we use the method of determining the fundamental parameters of LBVs 
developed by us using an example of AE And and Var A-1 and new LBVs that 
we discovered \citep{Sholukhova2015,Sarkisyan2020}. The method is based on modeling LBVs SEDs related to different states of a star, taking in to account intrinsic property of LBV stars - changing brightness in the optical spectral range with a nearly constant bolometric luminosity \citep{HumphreysDavidson1994}. The well known problem of $A_V - T$ parameter degeneracy may therefore be solved, since the constant bolometric luminosity and 
interstellar extinction for different LBV states allow us to constrain model parameters \citep{Sholukhova2011,Sholukhova2015}. As will be seen in Section~\ref{sect:phot}, our object J004341.84 shows the LBV-type variability, which enables us to use the method. 

\begin{figure}
	\includegraphics[angle=270,width=\columnwidth]{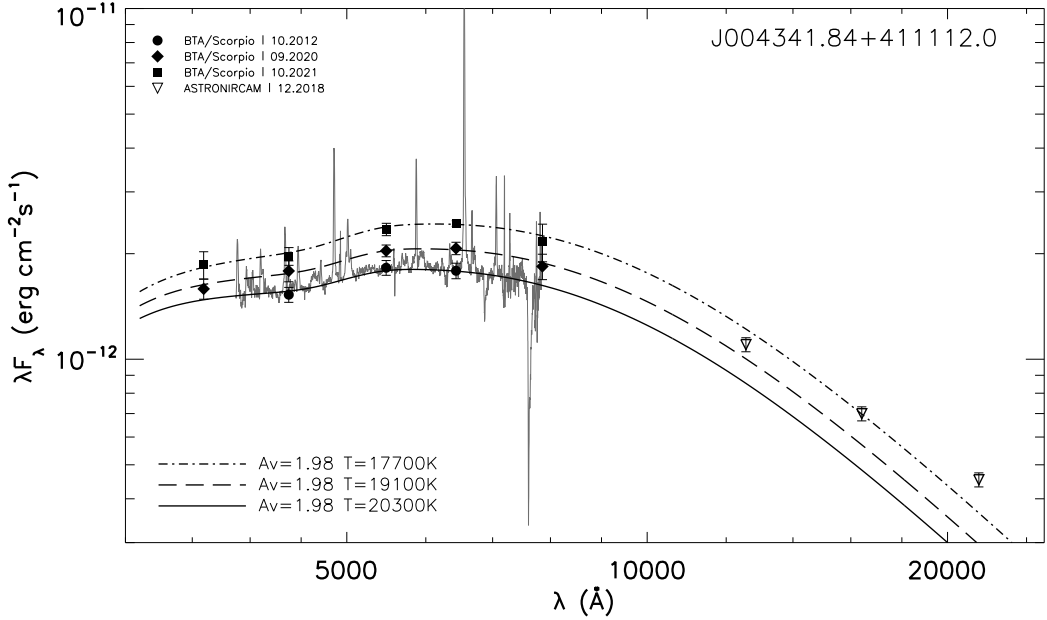}
	\caption{The spectra energy distributions of the J004341.84 in the optical and NIR ranges. 
		Only the phometric data for 2012, 2018, 2020, 2021 epochs and spectrum of 2012 are shown to avoid cluttering the figure. 
		The filled symbols designate the approximated data sets corrected for the emission lines and the spectra slope. 
		The model spectra are shown with the solid (for the 16.10.2012 epoch), dashed (for the 28.09.2020 epoch) and dash-dotted (for the 02.10.2021 epoch) lines. The legends in the figure indicate object name, symbols, instruments and dates for the photometry and  the best fitting parameters of temperature and reddening. }\label{fig:seds}
\end{figure}

We start with preliminary estimates of the stellar temperature ($T_{\text{sp}}$ 
in Table~\ref{tab:objpars}) using the visibility of lines in the spectra. Since the spectra of the object do not show significant variations at SCORPIO resolution, we define $T_{\text{sp}}$ range as 18000--22000\,K for all epochs. 
Next, we fit the photometric data points with black body spectrum using the constrained temperature and taking into account the dust extinction \citep{Fitzpatrick1999} with $R_V$ = 3.1. Following this way we fit five different sets of photometric data obtaned with SCORPIO (see Table~\ref{tab:phot}), setting bolometric luminosity and extinction to be constant for all sets, and estimate stellar parameters in corresponding states.

Figure~\ref{fig:seds} shows SED and its modelling result for our star. We used the optical photometry with the SCORPIO and NIR photometry with the ASTRONIRCAM (Table~\ref{tab:phot}) for SED construction. To avoid cluttering the figure, only the data for 2012, 2018, 2020, 2021 epochs and spectrum of 2012 are shown. We use the \textit{UBVRI} spectra range only for the approximation because the model does not take into account contribution from the bremsstrahlung radiation and the dust emission. We subtract emission lines contribution from the photometric points using our spectra and apply corrections to the effective band wavelengths for the spectra slope. Photometric points with those corrections are marked with the filled symbols in the Figure~\ref{fig:seds}. NIR photometric points were not used for fitting and are shown without any corrections by open symbols. The results of the SED approximation are shown with the solid, dashed and dash-dotted lines for the 2012, 2020 and 2021 epochs, respectively. The SED fitting results for remaining epochs presented in the Table~\ref{tab:objpars}, which shows best-fitting parameters and based on them estimated parameters: $A_V$, temperature, radius, and bolometric magnitude $M_{\text{bol}}$. We have to note that we accepted the distance to M31 of 752$\pm$27\,kpc \citep{Riess2012} in our calculation.

\begin{figure}
	\begin{center}
	\includegraphics[width=0.7\columnwidth]{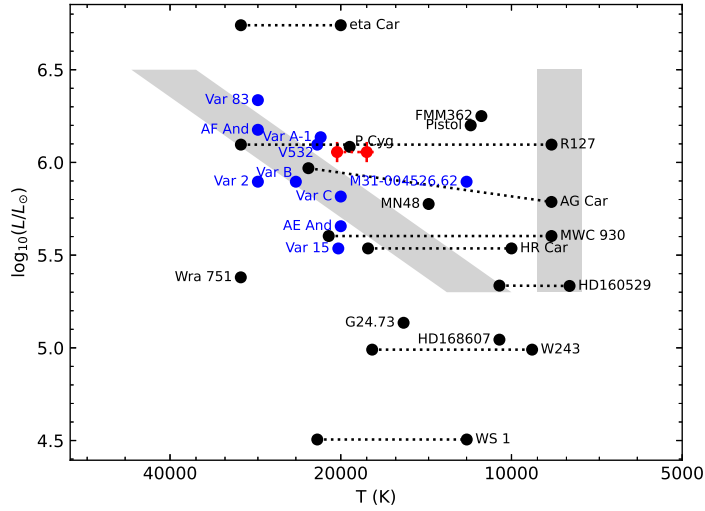}
	\caption{Position of the J004341.84 on the HR diagram. LBVs in M31 and M33 are shown with blue points accordingly to the data from the work of \citep{Humphreys2017a}. Positions of the Galactic LBVs and R127 taken from the similar figure of \citet{Smith2019} and shown in black. Position of the J004341.84 based on its SED modelling in current work shown in red. The grey boxes show the locations of the S Dor instability strip \citep{Wolf1989} and the constant temperature strip of LBVs in outburst.}\label{fig:hrd}
	\end{center}
\end{figure}

The method allows us to estimate the $A_V$ with appropriate accuracy, and therefore to estimate the bolometric luminosity of the star. Obtained values of extinction and luminosity are well match to estimates of \citet{Massey2006b}, and luminosity value correspond to that of known LBVs in M31 \citep{Humphreys2014}. Position of the J004341.84 on the Hertzsprung-Russell (HR) diagram shown on the figure~\ref{fig:hrd}. The object lie in immediate proximity to S Dor instability strip \citep{Wolf1989} and remarkably close to P Cyg and R127.  

The \textit{JHK} photometric points on SED clear show small NIR excess. However the shape of excess and its not particularly outstanding value in \textit{K} filter indicates that it relies on contribution of free-free (f-f) but not the warm dust emission. This excludes the possibility to classify the object as B[e] supergiant \citep{Humphreys2017b}. Furthermore, accordingly to the Spitzer/IRAC photometry \citep{Ravandi2016} the object have color index $[3.6]-[4.5]=0.212$ (based on isophotal magnitudes from catalog), which is correspond to LBVs region on diagram suggested by \citet[][figure 6 (b)]{Humphreys2017b}.       

\begin{table}
	\begin{center}
		\caption{SED fitting and estimated parameters for J004341.84. 
			The columns show the date of photometric data used for SED fitting and indicating the state of the object, 
			stellar photosphere temperature estimated from the spectrum, 
			temperature and reddening $A_{V}$ estimated from the SED fitting, 
			stellar radius, $M_V$ and $M_{\text{bol}}$. 
			Values of the columns $T_{\text{sp}}$, $A_{V}$ and $M_{\text{bol}}$ are common for all epochs (see text of Section~\ref{sect:sed}). \label{tab:objpars}}
		\begin{tabular}{c|c|c|c|c|c}
			\hline
			Date & $T_{\text{sp}}$ (K) & $T_{\text{SED}}$ (K) & $A_{V}$ (mag) & $R$ ($R_{\sun}$) &$M_{\text{bol}}$ (mag) \\
			\hline
			16.10.2012 &              & 20300$\pm$370  &               &  87$\pm$5  &                 \\ 
			17.01.2015 &              & 19300$\pm$370  &               &  96$\pm$3  &                 \\
			26.09.2016 & 17000--22000 & 18000$\pm$400  & 1.98$\pm$0.06 & 110$\pm$4  & -10.41$\pm$0.12 \\    
			25.10.2019 &              & 19000$\pm$350  &               &  99$\pm$4  &                 \\
			28.09.2020 &              & 19100$\pm$330  &               &  98$\pm$4  &                 \\ 
			02.10.2021 &              & 17700$\pm$300  &               &  114$\pm$4  &                 \\ 
			\hline
		\end{tabular}
	\end{center}
\end{table}

\section{Photomery and light curve}
\label{sect:phot}

\begin{figure}
	\includegraphics[width=0.9\columnwidth]{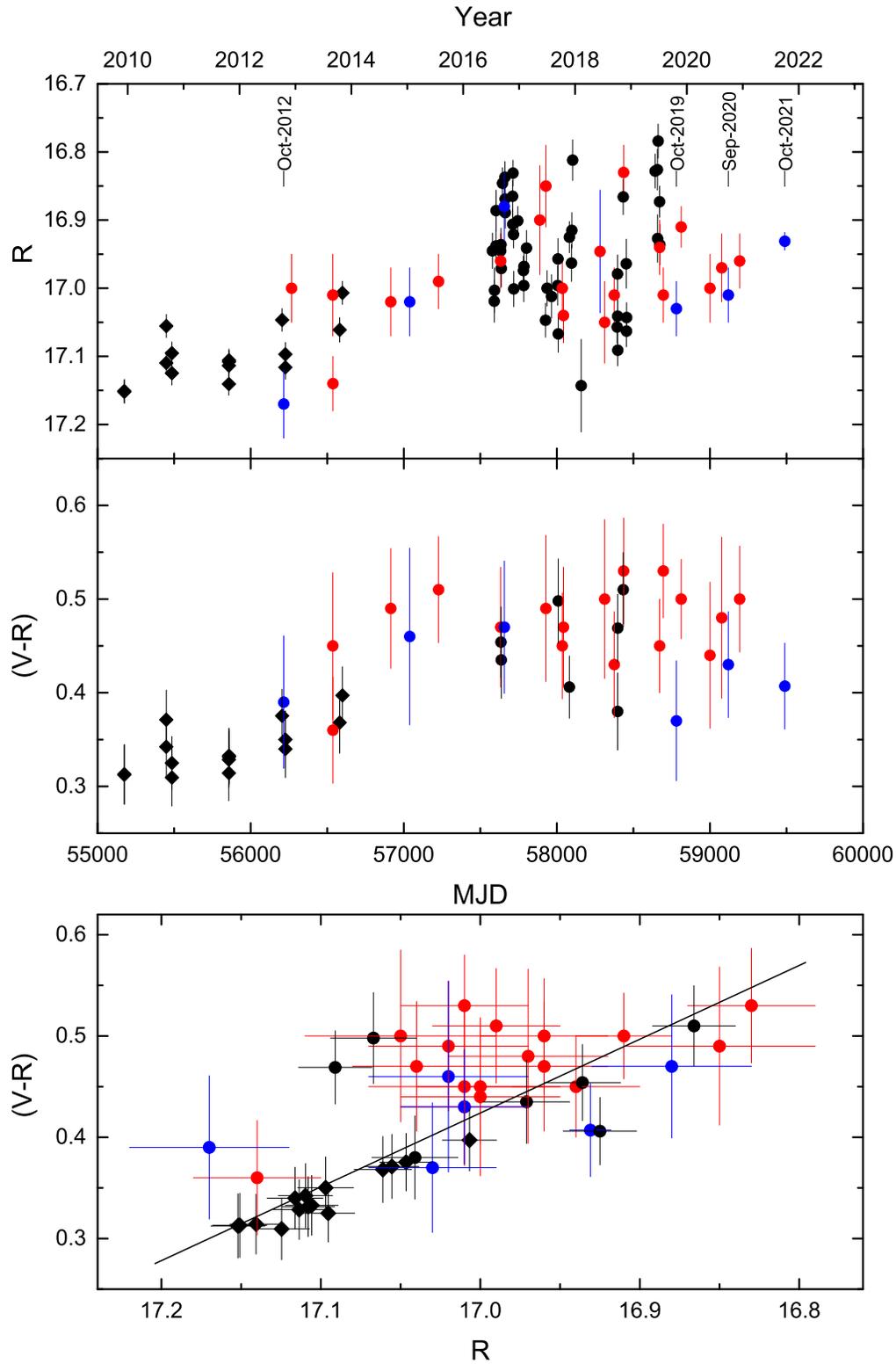}
	
	\caption{ (\textit{Top}) The \textit{R}-band light curve together with the (\textit{V}-\textit{R}) colour plotted against the Modified Julian Date (MJD) for the J004341.84. The arrows indicate the dates of our spectra. (\textit{Bottom}) The (\textit{V}-\textit{R}) vs. \textit{R} colour-magnitude diagram of the J004341.84.
	    Black diamonds -- data from Pan-STARRS \citep{Chambers2016}, 
		red circles -- data from \citep{Martin2017}, 
		black circles -- data from NAO Rozhen telescope,
		blue circles -- BTA photometry. 
		}\label{fig:lc}
\end{figure}

In Figure~\ref{fig:lc} we present a compiled light curve of J004341.84 in \textit{R} band constructed from our and archive data for the period from 2010 to 2021. In addition to our data we also use photometric data of Panoramic Survey Telescope and Rapid Response System (Pan-STARRS, \citet{Chambers2016}) and data from the work of \citet{Martin2017}. To adopt data from Pan-STARRS archive we use equations for photometric transformations between Pan-STARRS and other systems from the work of \citet{Tonry2012}. The vertical lines indicate the dates of obtaining the spectra on the BTA telescope. 

To confirm object variability we use the robust median statistic as the variability index \citep{Enoch2003,Rose2007}: 
 $ \bar{\eta}=\frac{1}{N-1}\sum _{i=1}^{N}\left|\frac{m_{i}-\tilde{m}}{\sigma _{i}}\right|$.
For a constant star, the value would be close to unity whereas stars with values above 1 are counted as photometrically variable objects. The calculated value for our light curve is $\bar{\eta}=3.59$.
The object shows the variability range $\Delta$\textit{R}\,$\approx$\,0.4\,mag. Despite the brightness variation is not outstanding, light curve clear shows brightness increasing trend over the last decade with $\Delta$\textit{R}\,$>$\,0.1\,mag and it is quite enough to classify the candidate as a LBV in quiescent phase \citep{vanGenderen2001}.

Moreover, the colour variations of the object strongly supports the LBV classification. As we see from the Figure~\ref{fig:lc}, the (\textit{V}-\textit{R}) colour becomes redder with time and, more importantly, with brightness increase, corresponding to an actual temperature decrease. This is fully consistent with the LBV behavior when a brightness increase as the temperature decreases due to radius inflation, while the object's bolometric luminosity stays roughly constant \citep{HumphreysDavidson1994}. Since this feature is distinctive to S Doradus variables \citep{Lamers1998,vanGenderen2001}, the light curve of the J004341.84 and its color variations confirm LBV classification of the object.

\section{Conclusions}

We study LBV candidate J004341.84 using our spectra and photometry as well archival photometric data. 
In order to confirm LBV status of the star we analysed light curve of this object in \textit{R} filter and estimated the stellar parameters from spectral energy distribution of the object.

\citet{Massey2007} and \citet{Humphreys2014} classify this LBV candidate as an Of/late-WN star. 
A connection between the Of/late-WN and LBV classes has been established by the transition one of the prototypes of Of/late-WN stars in LMC, R127 (HDE 269858), to LBV phase \citep{Stahl1983,Stahl1986,Wolf1988}. Conversely, galactic LBV AG Carinae showed Of/late-WN spectrum in its light minimum state \citep{Stahl1986a}. Subsequently, relationship of two classes was extended by a number of Of/late-WN stars turned out to be LBV and LBVs or LBV candidates with Of/late-WN spectra: He 3-591 in our Galaxy; HD 5980 in SMC; R127, R71, HDE 269582 in LMC; AF And, Var 15 in M31; and Var B, Var 2, MCA-1B, V532 in M33. It has been even assumed that Of/late-WN stars are quiescent LBVs and perhaps all LBVs at hot state are Of/late-WN objects \citep{Bohannan1989}. During our spectral monitoring of the object we have noted the variation of the one of distinctive lines for Of/late-WN stars \ion{He}{II} 4686: it significantly weakened from 2012 to 2019 and then strengthened again from 2019 to 2020 and 2021. Similar behavior with disappearing \ion{He}{II} 4686 line were demonstrated by two Of/late-WN stars R127 \citep{Stahl1983} and HDE 269582 \citep{Stahl1986a}, that subsequently have been classified as LBVs \citep{Walborn2017}. This may point to typical LBV instability of the object. 

\citet{Massey2006a} pointed that the object is an analog of the well known LBV star P Cygni. 
We also notice P Cygni profiles in the hydrogen, \ion{Fe}{iii} and \ion{He}{i} lines and close matching of the spectrum to those of P Cygni and P Cygni-like LBV candidate MN112. This is indicative of common nature of the objects and supports LBV classification of the J004341.84.      

We continue to apply new method of LBV SED modeling \citep{Sholukhova2015,Sarkisyan2020} to studying our LBV candidate. 
Our fitting of SEDs of several stellar states with constant extinction and bolometric 
luminosity yields $A_V=1.98\pm0.07$ and $M_{\text{bol}}=-10.41\pm0.12$\,mag. The obtained estimates of extinction and luminosity are well match to those ones of \citet{Massey2006b}, and luminosity value correspond to that of known LBVs in M31 \citep{Humphreys2014}. Estimated luminosity and temperature of the J004341.84 put it in immediate proximity to S Dor instability strip \citep{Wolf1989} and remarkably close to P Cyg and R127 on HR diagram. 

Light curve of J004341.84 for the period from 2010 to 2020 displays variations of the order of 0.4\,mag, with an overall brightness increasing trend $\Delta$\textit{R}\,$>$\,0.1\,mag. In full accordance to S Dor variables behavior \citep{HumphreysDavidson1994,Lamers1998,vanGenderen2001} the increase in brightness of the object corresponds to a redder colour and therefore a cooler temperature.

Taking into account S Dor-type variability of the object and similarity of its spectrum and estimated luminosity to those of known LBVs, J004341.84 can be classified as a LBV star.

\section*{Acknowledgements}

The reported study was founded by RFBR and NSFB according to the research project N 19-52-18007.
Observations with the SAO RAS telescopes are supported by the Ministry of Science and Higher Education of the Russian Federation (including agreement No05.619.21.0016, project ID RFMEFI61919X0016). S. F., P. N. and A. Valcheva acknowledge partial support from DN18-10/2017 grant from the NSF of Bulgaria. This work was performed with the equipment purchased from the funds of the Program of Development of Moscow University. The authors also thank the anonymous referee for their careful review and helpful suggestions that improved the manuscript.

\bibliographystyle{raa}
\bibliography{ms2021-0239}

\end{document}